# Exploring A Multi-Scale Method for Molecular Simulations in Continuum Solvent Model: Explicit Simulation of Continuum Solvent As An Incompressible Fluid

Li Xiao[1,2] and Ray Luo[1,2,3]

1. Departments of Biomedical Engineering, 2. Molecular Biology and Biochemistry, and 3. Chemical Engineering and Materials Science, University of California, Irvine, CA 92697

A multi-scale framework was recently proposed for more realistic molecular dynamics simulations in continuum solvent models by coupling a molecular mechanics treatment of solute with a fluid mechanics treatment of solvent, where we formulated the physical model and developed a numerical fluid dynamics integrator. In this study, we incorporated the fluid dynamics integrator with the Amber simulation engine to conduct atomistic simulations of biomolecules. At this stage of the development, only nonelectrostatic interactions, i.e., van del Waals and hydrophobic interactions are included in the multi-scale model. Nevertheless numerical challenges exist in accurately interpolating the highly nonlinear van del Waals term when solving the finite-difference fluid dynamics equations. We were able to bypass the challenge rigorously by merging the van del Waals potential and pressure together when solving the fluid dynamics equations and by considering its contribution in the free-boundary condition analytically. The multi-scale simulation engine was first validated by reproducing the solute-solvent interface of a single atom with analytical solution. Next, we performed the relaxation simulation of a restrained symmetrical monomer and observed a symmetrical solvent interface at equilibrium with detailed surface features resembling those found on the solvent excluded surface. Four typical small molecular complexes were then tested, both volume and force balancing analysis showing that these simple complexes can reach equilibrium within the simulation time window. Finally, we studied the quality of the multi-scale solute-solvent interfaces for the four tested dimer complexes and found they agree well with the boundaries as sampled in the explicit water simulations.

# 1. Introduction

Atomistic simulation has become an important tool for studying the structures, dynamics, and functions of biomolecular systems. Nevertheless efficient atomistic simulation of large and complex biomolecular systems is still one of the remaining challenges in computational molecular biology. The computational challenges in atomistic simulation of biomolecular systems are direct consequences of their high dimensionalities. Indeed biomolecules are highly complex molecular machines with thousands to millions of atoms. What further complicates the picture is the need to realistically treat the interactions between biomolecules and their surrounding water molecules that are ubiquitous and paramount important for their structures, dynamics, and functions.

To appreciate these challenges, it is instructive to highlight the two bottlenecks in biomolecular simulations: (1) the cost of each energy evaluation that is determined by the number of particles in a mathematical model; and (2) the number of time steps of dynamics that it takes for sufficient coverage of different conformations. Hundreds of millions of time steps are routinely required in biomolecular simulations to draw statistically significant conclusions. It is often the case that more particles need more time steps for sufficient coverage. Thus the overall simulation cost usually scale exponentially with the number of particles in the mathematical model used in a simulation. Indeed many fundamental and interesting biomolecular processes remain largely inaccessible to atomistic simulations when system sizes exceed more than a few hundred residues.

Since most particles in biomolecular simulations are to represent water molecules solvating the target biomolecules, an implicit treatment of water molecules allows greatly increased simulation efficiency. Indeed, implicit solvation offers a unique opportunity for more efficient simulations without the loss of atomic-level resolution for biomolecules [1-17]. Advance in implicit solvation, coupled with developments in sampling algorithms, classical force fields, and quantum approximations, will prove useful to the larger biomedical community in a broad range of studies of biomolecular structures, dynamics and functions.

One class of implicit solvent models, the classical Poisson-Boltzmann (PB) solvent model, has become widely accepted in biomolecular applications after over 30 years of basic research and development. Efficient numerical PBE-based solvent models have been widely used to study biological processes including predicting p$K$a values [18-21], computing solvation and binding free energies [22-31], and protein folding [32-42]. However, challenges remain to achieve more consistent, accurate, and robust analysis of biomolecules [43-60]. The existing dielectric model based on molecular solvent excluded surface is a major hurdle for applications of the Poisson-Boltzmann solvent models. This dielectric model is ad-hoc, expensive, and numerically unstable due to its treatment of atoms as hard spheres in molecular simulations.

In our previous study [61], we explored a multi-scale simulation strategy to explicitly simulate the continuum solvent/solute interface with the solvent fluid dynamics that is coupled to the solute molecular dynamics. This strategy (1) allows a self-consistent treatment of the solvation interactions, i.e. the dielectric interface automatically adjusts to local conformational and energetic fluctuations and is guaranteed to be at the system free energy minimum upon equilibrium; (2) allows a "soft" and more physical dielectric interface for stable dynamics; (3) eliminates atom-specific cavity radii that must be defined, dramatically reducing the freely adjustable parameters of the continuum solvent treatment; (4) eliminates the expensive molecular surface reconstruction step during dynamics; and (5) eliminates the difficult and expensive molecular surface-to-atom mapping of dielectric boundary forces and hydrophobic boundary forces, and applies these surface forces to the continuum solvent instead. In addition, a 3D numerical algorithm was developed to simulate the implicit solvent via the Navier-Stokes equation [61, 62]. It should be pointed out that the use of Navier-Stokes equation, instead of Stokes equation that is sufficient for biomolecular processes of interest, is necessary for the lack of a predefined solute-solvent interface, or in a "free boundary" problem [61, 62]. Our numerical algorithm was validated with multiple model test cases, demonstrating its effectiveness and numerical stability, with observed accuracy consistent with the designed numerical algorithm.

In this study, we intended to explore the feasibility of incorporating the fluid dynamics algorithm into the Amber molecular mechanics simulation engine [63-65] to assess the feasibility and quality of the new multi-scale model for potential applications to biomolecular simulations. At the current stage, we are mainly interested in equilibrium properties of the biomolecular solute, and solvent hydrodynamics is not our consideration. Thus certain alterations of the original model can be utilized to artificially accelerate the solvent relaxation process so that the precious computing resources can be focused on sampling of solute conformations.

## 2. Theoretical Model

In the following we first review our physical model for easy understanding of the overall approach. Next we briefly go over the fluid dynamics algorithm and procedure with a focus on what has been revised from our previous study to adapt the method to atomistic molecular simulations. Finally computational details are presented for the numerical tests of specific molecular systems.

### 2.1 Physical model

Our basic model is derived from the Hamiltonian equation. A Hamiltonian for the entire system is thus defined first. Its degrees of freedom are atomic positions ( $\mathbf{x}$ ) and their velocities ( $\mathbf{v}$ ) for the solute molecular dynamics (MD) region; and fluid element displacements ($\mathbf{y}$) and their velocities ($\mathbf{u}$) for the solvent fluid dynamics (FD) region. For the MD region, all-atom molecular mechanics will be used. Molecular mechanics usually adopts a relatively simple potential energy function, or force field, for efficient computation. Many potential energy functions have been developed for biomolecular applications, such as Amber [66-71], CHARMM [72-74] and OPLS [75-77]. For FD region, an incompressible viscous fluid model is adopted.

The Hamiltonian is defined as

$$H = H_{MD}(\mathbf{x}, \alpha_{\mathbf{x}}) + H_{FD}(\mathbf{y}, \alpha_{\mathbf{y}}) + H_{MD/FD}(\mathbf{x}, \alpha_{\mathbf{x}}; \mathbf{y}, \alpha_{\mathbf{y}}), \tag{1}$$

where $\alpha_x$ is the momentum of MD region and $\alpha_y$ is the momentum of FD region. $H_{MD}$ is the Hamiltonian of for the MD region modeled by molecular mechanics $H_{MD} = U + K$, where $U$ is the force field potential energy and $K$ is the kinetic energy. $H_{FD}$ represents the Hamiltonian for the incompressible solvent fluid. $H_{MD/FD} = U_{ele} + U_{vdw} + U_{hse}$, consists of three terms. $U_{ele}$ is the Poisson-Boltzmann electrostatic solvation energy [78-81]. The nonelectrostatic solvation energy is modeled as two components: the van del Waals component $U_{vdw}$ and the hard sphere entropy/cavity component $U_{hse}$ [82-86]. Here $U_{ele}$ is defined as

$$U_{ele} = \int \left( \rho^f \varphi - \frac{1}{8\pi} \mathbf{D} \cdot \mathbf{E} - \Delta \Pi \lambda \right) dv$$
$$\Delta \Pi = kT \sum_i c_i (e^{-q_i \varphi / kT} - 1) \tag{2}$$

and $U_{vdw}$ and $U_{hse}$ are defined as

$$U_{vdw} = \sum_{a=1}^{N_s} \int \rho_{aw}(\mathbf{r}_{aw}) u_{LJ}(\mathbf{r}_{aw}) d\mathbf{r}_{aw}$$
$$U_{hse} = \gamma \bullet SAS + c \tag{3}$$

Here the sum is over all solute atoms ($N_s$), and the integration is over the solvent-occupied volume. $\rho_{aw}(\mathbf{r}_{aw})$ is a solvent distribution function around solute 'a' at a given solute-solvent distance. $u_{LJ}(r) = \dfrac{A}{r^{12}} - \dfrac{B}{r^6}$ is the force field Lennard Jones potential given the coefficient A, B for each atom. $\gamma$ is the surface tension and $c$ is an offset constant.

Now we proceed to derive the dynamics equation by first setting $\beta = (\mathbf{x}, \mathbf{y})$ as the position vector of the system and $\alpha = (\alpha_\mathbf{x}, \alpha_\mathbf{y})$ as the momentum vector of the system. The familiar Newtonian dynamics can be derived from the Hamilton's equation

$$\dot{\alpha} = -\frac{\partial H}{\partial \beta}. \tag{4}$$

Here we have adopted the convention that $\alpha$ and $\beta$ represent the moment and position vectors of each particle/element, respectively.

In the molecule dynamics region, the equation of motion for an atom at position vector $x$ can be expressed symbolically as

$$\dot{\alpha_x} = -\frac{\partial H_{MD}}{\partial x} - \frac{\partial H_{MD/FD}}{\partial x}. \tag{5}$$

$-\dfrac{\partial H_{MD}}{\partial x}$ represents the usual force field terms in molecule dynamics simulations. The coupling Hamiltonian has three terms, $U_{ele} + U_{vdw} + U_{hse}$. Since $U_{hse}$ does not depend on atomic positions, the coupling force terms that the atoms feel are only those of electrostatics and van del Waals in nature, i.e., $-\dfrac{\partial H_{MD/FD}}{\partial x} = -\dfrac{\partial U_{ele}}{\partial x} - \dfrac{\partial U_{vdw}}{\partial x}$. It is interesting to note that the electrostatic forces $-\dfrac{\partial U_{ele}}{\partial x}$ are simply the $q\mathbf{E}$ forces, where $q$'s are the "free" charges, i.e. atomic point charges in a force field model [87]. $-\dfrac{\partial U_{vdw}}{\partial x}$ are the van del Waals forces from the solvent molecules modeled as continuum [84] .

In the fluid dynamics region, consider a small fluid volume element at position $y$, with volume $V$ and velocity $\mathbf{u}$. The equation of motion of the fluid element is

$$\dot{\alpha}_y = -\frac{\partial H_{FD}}{\partial y} - \frac{\partial H_{MD/FD}}{\partial y}. \tag{6}$$

As shown below, the variational principle will be applied on this element. The partial derivative can also be written in the variational form as

$$\dot{\alpha}_y = -\frac{\delta H_{FD}}{\delta y}\Big|_{\mathbf{u}} - \frac{\delta H_{MD/FD}}{\delta y}\Big|_{\mathbf{u}}. \tag{7}$$

Here the subscript $\mathbf{u}$ denotes that it is fixed during the variation. Notice here $y = y(b,t)$ is the Lagrangian coordinate of the volume element, which is fixed on the fluid element and $b$ is introduced here to denote the actual spatial position [88].

Let us first focus on the variation of $H_{FD}$, which has the form of

$$H_{FD} = \int \left[ \frac{1}{2} \rho \mathbf{u}^2 + U_{\text{int}}(\rho,s) \right] dV, \tag{8}$$

where $U_{\text{int}}$ is the internal energy density, $\rho$ represents the fluid density, and $s$ is entropy. For a small fluid volume element at position $y$ and volume $V$, we impose a variation $\delta y$ on the element within time $dt$, and proceed to compute the variation of $H_{FD}$. The process is assumed to be very rapid, i.e., $\dfrac{\delta y}{dt} \gg 0$. Since the fluid is incompressible, $\nabla \bullet \delta y = 0$, the work done to the environment is

$$dW = \int p\,\delta y \bullet d\mathbf{S} = \int \nabla \bullet (p\,\delta y) dV = \int \nabla p \bullet \delta y dV + \int p \nabla \bullet \delta y dV = \int \nabla p \bullet \delta y dV, \tag{9}$$

where $p$ is pressure and vector $d\mathbf{S}$ denotes the surface element of the element with the direction along the normal direction of the surface. Given that the first and second laws of thermodynamics still hold, the internal energy variation can be expressed as

$$\int \delta U_{\text{int}}\, dV = \int (T\,\delta s)dV - dW = \int (T\,\delta s)dV - \int \nabla p \bullet \delta y dV. \tag{10}$$

The entropy constraint also gives

$$\int (T\,\delta s)\, dV = \int \mathbf{f}_{vis} \bullet \delta y\, dV - dt \int \nabla \bullet \mathbf{q}\, dV, \tag{11}$$

where $\mathbf{q}$ is the heat flux and $\mathbf{f}_{vis}$ is the viscous force density. Substitution of eqn (11) into eqn (10) and the fact that the term involving $dt$ can be ignored as $\dfrac{\delta y}{dt} \gg 0$ given the variation of the internal energy as

$$\int \delta U_{\text{int}}\, dV = \int (\mathbf{f}_{vis} \bullet \delta y)dV - \int \nabla p \bullet \delta y dV = \int (\mathbf{f} \bullet \delta y)dV, \tag{12}$$

where the total force density $\mathbf{f} = \mathbf{f}_{vis} - \nabla p = \dfrac{\partial \sigma_{ij}}{\partial y_j}$ is introduced and

$\sigma_{ij} = -p\delta_{ij} + \mu\left(\dfrac{\partial(\mathbf{u} \cdot y_i)}{\partial y_j} + \dfrac{\partial(\mathbf{u} \cdot y_j)}{\partial y_i}\right)$ is the stress tensor and $\mu$ is the fluid viscosity constant [89].

of the fluid [89]. Given the assumption that the force density is uniform within the volume element, substitution of eqn (8) and eqn (12) into the variation of $H_{FD}$ gives

$$-\dfrac{\delta H_{FD}}{\delta y}\Big|_{\mathbf{u}} = -\dfrac{\int \delta U_{\text{int}}\, dV}{\delta y} = -\mathbf{f}V = -\dfrac{\partial \sigma_{ij}}{\partial y_j}V. \tag{13}$$

The variation of $H_{MD/FD}$ is presented next. In Poisson-Boltzmann systems with mobile ions, there is an ionic force term at the Stern layer [87], but it is usually much smaller than other force terms, and is often ignored. If it were not ignored, the ionic force would act upon relevant volume elements. $U_{hse}$ only depends on the interface boundary so that it does not change under the variation of the volume element. Thus the only significant derivative of $H_{MD/FD}$ is the van del Waals force, which can be treated as the "external force" density ($\mathbf{F}$) on the fluid element, i.e.

$$-\frac{\delta H_{MD/FD}}{\delta y}\Big|_{\mathbf{u}} = -\frac{\delta U_{vdw}}{\delta y} = \mathbf{F}V. \tag{14}$$

Finally, the change of momentum of the fluid volume element is

$$\dot{\alpha}_y = \frac{d(\rho V\mathbf{u})}{dt} = \rho V\frac{\partial \mathbf{u}}{\partial t} + \sum_i \rho V\frac{\partial \mathbf{u}}{\partial a_i}\frac{\partial a_i}{\partial t} = \rho V\frac{\partial \mathbf{u}}{\partial t} + \rho V(\mathbf{u}\bullet\nabla)\mathbf{u}. \tag{15}$$

Combination of eqns (6) and (13) –(15) gives

$$\rho V\frac{\partial \mathbf{u}}{\partial t} + \rho V(\mathbf{u}\bullet\nabla)\mathbf{u} = \frac{\partial \sigma_{ij}}{\partial y_j}V + \mathbf{F}V$$

$$\Rightarrow \rho\frac{\partial \mathbf{u}}{\partial t} + \rho(\mathbf{u}\bullet\nabla)\mathbf{u} = \frac{\partial \sigma_{ij}}{\partial y_j} + \mathbf{F}. \tag{16}$$

Including the conservation of volume/mass for the given volume element, i.e. $\nabla\bullet\mathbf{u} = 0$, the incompressible Navier-Stokes equation can be expressed as

$$\rho\left(\frac{\partial \mathbf{u}}{\partial t} + (\mathbf{u}\cdot\nabla)\mathbf{u}\right) = -\nabla p + \mu\Delta\mathbf{u} + \mathbf{F}$$

$$\nabla\cdot\mathbf{u} = 0 \tag{17}$$

## 2.2 Derivation of interface conditions

To obtain the interface conditions, an infinitely small fluid disk element $\varepsilon$ is introduced with small area $A$ and thickness $h$, and with $h \ll \sqrt{A}$. The disk surfaces are parallel to the boundary interface and one side of the surface is in the molecule dynamics region. Given a variation of the disk position with $\delta\mathbf{r}_\varepsilon$,

$$\dot{\alpha}_\varepsilon = -\frac{\delta H_{FD}}{\delta\mathbf{r}_\varepsilon} - \frac{\delta H_{FD/MD}}{\delta\mathbf{r}_\varepsilon}. \tag{18}$$

On the left-hand side

$$\dot{\alpha_\varepsilon} = \rho A h \frac{d\mathbf{u}}{dt}. \tag{19}$$

On the right-hand side, we first introduce the local coordinate system, which consists of one normal direction ($\mathbf{n}$) and two tangential directions ($\mathbf{t}, \tau$) at a certain point on the interface, i.e.,

$$\begin{cases} \mathbf{n} = \cos\alpha_1\mathbf{i} + \cos\alpha_2\mathbf{j} + \cos\alpha_3\mathbf{k} \\ \mathbf{t} = \cos\beta_1\mathbf{i} + \cos\beta_2\mathbf{j} + \cos\beta_3\mathbf{k} \\ \tau = \cos\gamma_1\mathbf{i} + \cos\gamma_2\mathbf{j} + \cos\gamma_3\mathbf{k} \end{cases}. \tag{20}$$

Stress $-\dfrac{\delta\mathbf{H}_{FD}}{\delta\mathbf{r}_\varepsilon}$ only exerts on the disk surface in the fluid region, so that

$$-\frac{\delta\mathbf{H}_{FD}}{\delta\mathbf{r}_\varepsilon} = \sigma_{ij}\bullet\mathbf{n}A = \left[(-p + 2\mu\frac{\partial\mathbf{u}}{\partial\mathbf{n}}\bullet\mathbf{n})\mathbf{n} + \mu(\frac{\partial(\mathbf{u}\cdot\mathbf{n})}{\partial\mathbf{t}} + \frac{\partial(\mathbf{u}\cdot\mathbf{t})}{\partial\mathbf{n}})\mathbf{t} + \mu(\frac{\partial(\mathbf{u}\cdot\mathbf{n})}{\partial\tau} + \frac{\partial(\mathbf{u}\cdot\tau)}{\partial\mathbf{n}})\tau\right]A. \tag{21}$$

The last term of eqn (18) can be worked out as

$$-\frac{\delta H_{MD/FD}}{\delta\mathbf{r}_\varepsilon} = -\frac{\delta U_{ele}}{\delta\mathbf{r}_\varepsilon} - \frac{\delta U_{hse}}{\delta\mathbf{r}_\varepsilon} - \frac{\delta U_{vdw}}{\delta\mathbf{r}_\varepsilon}, \tag{22}$$

where $-\dfrac{\partial U_{ele}}{\partial\mathbf{r}_\varepsilon} = \mathbf{f}_{dielec}A = \dfrac{1}{2}\sigma^{pol}\dfrac{\mathbf{D}_i\bullet\mathbf{D}_o}{D_{o\mathbf{n}}}\mathbf{n}A$ is the dielectric boundary electrostatic force [87]. The term $-\dfrac{\partial U_{hse}}{\partial\mathbf{r}_\varepsilon} = -\gamma\kappa\mathbf{n}A$ is the pressure and surface tension from the hard sphere entropy, aka the hydrophobic term [84], where $\kappa$ is the curvature. The van del Waals force, $\dfrac{\delta U_{vdw}}{\delta\mathbf{r}_\varepsilon}$ [84], proportional to the volume of element $Ah$, can be ignored when comparing to the electrostatic forces and surface tension as $h$ is infinitely small. Combining eqns (18) and (19) and the terms calculated above, the interface conditions can be summarized as

$$-p + p_g - \gamma\kappa + f_{dielec} + 2\mu\frac{\partial \mathbf{u}}{\partial \mathbf{n}} \bullet \mathbf{n} = 0$$

$$\frac{\partial(\mathbf{u}\cdot\mathbf{n})}{\partial \mathbf{t}} + \frac{\partial(\mathbf{u}\cdot\mathbf{t})}{\partial \mathbf{n}} = 0 \qquad \text{on } \partial\Omega \qquad (23)$$

$$\frac{\partial(\mathbf{u}\cdot\mathbf{n})}{\partial \tau} + \frac{\partial(\mathbf{u}\cdot\mathbf{\tau})}{\partial \mathbf{n}} = 0,$$

## 3. Numerical Algorithms

We explored to implement the multi-scale model in numerical simulations with a strategy similar to those of the classical Car-Parrinello molecular dynamics (CPMD) model [90], which can be regarded as a multi-scale model via coupling equations of motion for ions and electrons in two different mechanics. In CPMD electrons are treated as active degree of freedom, via fictitious dynamics variable, and the fictitious electron dynamics is coupled with ionic dynamics in the Berendsen heat bath to approach the Born-Oppenheimer surface. The CPMD model results in a conservative ionic dynamics that is extremely close to the Born-Oppenheimer surface.

Our approach is to couple equations of motion for solute atoms and continuum solvent. The solvent part is also treated by the fictitious dynamics variable, and since our model is based on a finite-difference method, it is the fluid element. The fictitious fluid dynamics is modeled by the incompressible Navier-Stokes (NS) equation. The fictitious fluid dynamics model is coupled with all-atom molecular dynamics model in the Berendsen heat bath to approach the surface provided by all-atom MD simulations at a preset temperature. In doing so, the changes to the existing molecular mechanics simulation engine can be kept at the minimal and there is a very clear boundary between the FD and MD simulation routines, facilitating the development of the new model into a viable simulation engine for future biomolecular applications.

### 3.1 FD time integration

Our previous work has addressed the mathematical issues in solving fluid dynamics equations numerically [61, 62]. After setting the water density to unity, the velocity can be solved by the second-order semi-implicit backward Euler method as

$$\frac{3\mathbf{u}^{k+1} - 4\mathbf{u}^k + \mathbf{u}^{k-1}}{2\Delta t} + (\mathbf{u} \cdot \nabla \mathbf{u})^{k+1} = -\nabla p^{k+1} + \mu \Delta \mathbf{u}^{k+1} + \mathbf{F}^{k+1}, outside$$

$$\frac{3\mathbf{u}^{k+1} - 4\mathbf{u}^k + \mathbf{u}^{k-1}}{2\Delta t} = \mu \Delta \mathbf{u}^{k+1}, inside$$

(24)

where

$$p^{k+1} = 2p^k - p^{k-1}$$
$$(\mathbf{u} \cdot \nabla \mathbf{u})^{k+1} = 2(\mathbf{u} \cdot \nabla \mathbf{u})^k - (\mathbf{u} \cdot \nabla \mathbf{u})^{k-1}.$$

(25)

The pressure is solved by:

$$\Delta p^{k+1} = -\nabla \bullet ((\mathbf{u}^{k+1} \bullet \nabla)\mathbf{u}^{k+1}) + \nabla \bullet \mathbf{F}^{k+1}.$$

(26)

A new issue facing the application of the FD model to molecular simulation is the presence of van der Waals force ($\mathbf{F}$), which has a large gradient nearby the interface because it is too close to the solute atom centers. The large gradient is almost always challenging to address with a finite-difference type of method. In this study, we overcome the issue by introducing a variable $p'$, where $p' = p + \Gamma$ with $\nabla \Gamma = -\mathbf{F}$ obtained analytically. Therefore, we can solve $\nabla p' = \mu \Delta \mathbf{u}$ without computing the numerical gradient of the van der Waals potential. Specifically given $p'^{(k+1)} = p^{k+1} + \Gamma^{k+1}$, the equivalent form of eqn (26) to be solved numerically as

$$\Delta p'^{(k+1)} = -\nabla \bullet ((\mathbf{u}^{k+1} \bullet \nabla)\mathbf{u}^{k+1})$$

(27)

Accordingly, eqn (24) is updated as

$$\frac{3\mathbf{u}^{k+1} - 4\mathbf{u}^{k} + \mathbf{u}^{k-1}}{2\Delta t} + (\mathbf{u}\cdot\nabla\mathbf{u})^{k+1} = -\nabla p^{'(k+1)} + \mu\Delta\mathbf{u}^{k+1}, outside$$

$$\frac{3\mathbf{u}^{k+1} - 4\mathbf{u}^{k} + \mathbf{u}^{k-1}}{2\Delta t} = \mu\Delta\mathbf{u}^{k+1}, inside$$

(28)

where $p^{'(k+1)}$ is taken as

$$p^{'(k+1)} = 2p^{'(k)} - p^{'(k-1)}$$

(29)

Finally the interface boundary condition eqn (23) becomes

$$-p^{'} + \Gamma + p_{g} - \gamma\kappa + f_{dielec} + 2\mu\frac{\partial\mathbf{u}}{\partial\mathbf{n}}\bullet\mathbf{n} = 0$$

$$\frac{\partial(\mathbf{u}\cdot\mathbf{n})}{\partial\mathbf{t}} + \frac{\partial(\mathbf{u}\cdot\mathbf{t})}{\partial\mathbf{n}} = 0$$

$$\frac{\partial(\mathbf{u}\cdot\mathbf{n})}{\partial\tau} + \frac{\partial(\mathbf{u}\cdot\tau)}{\partial\mathbf{n}} = 0,$$

(30)

At each time step, the $p^{'}$ is interpolated with the one-side least square fitting method.[91] $\Gamma$ is computed analytically for each interface point where the interface boundary condition eqn (30) is enforced. When doing so, we can completely avoid finite-difference operations involving van del Waals energy and forces.

As presented in our previous works, the remaining major mathematical challenge in solving these coupled partial differential equations is the presence of the free boundary condition eqn (23) that allows the solute-solvent interface to equilibrate according to our physical model. To enforce the free boundary condition when solving pressure or velocity, we utilized the jump conditions of $\mathbf{u}_n$ and $p_n$ as the augmented variables, respectively [61, 62, 92]. The considerations of augmented variables lead to extra correction terms on the right-hand side in eqns (24) and (26). After the correction, each velocity component solver is equivalent to a Helmholtz equation. Once the velocity is updated, the pressure solver is simplified to a Poisson equation. In this implementation, we utilized the MICCG numerical solver to solve these linear differential equations [93-96].

When solving the linear systems, the fluid domain is contained in a rectangular box, the conditions at the outer boundary of the rectangular $R$ are

$$
\begin{aligned}
\mathbf{u}|_{x=x_{\min}} = 0, \qquad \mathbf{u}|_{x=x_{\max}} = 0 \\
\left.\frac{\partial \mathbf{u}}{\partial y}\right|_{y=y_{\min}} = 0, \qquad \left.\frac{\partial \mathbf{u}}{\partial y}\right|_{y=y_{\max}} = 0 \\
\mathbf{u}|_{z=z_{\min}} = 0, \qquad \mathbf{u}|_{z=z_{\max}} = 0 \\
p = 0,
\end{aligned}
\qquad \text{on } \partial R, \qquad (31)
$$

which represents a pipe flow in the $y$ direction. The use of the boundary condition allows the mass conservation law to be preserved since the incompressible solvent fluid can go in and out of the simulation box freely.

### 3.2 FD/MD interface update

Once the fluid velocity field is known, the next step is to use it to update the solute/solvent interface. The equivalent step in the solute region is to update particle positions based on particle velocities. Numerically we use the level set method based on the finite-difference method [97-99]. In the level set method, a scalar function, i.e. the level set function, is used to represent the moving interface implicitly. The interface is located where the level set function is zero ($d$=0), i.e. the zero level set $\Gamma(t) = \{\mathbf{y} : d(\mathbf{y}, t) = 0\}$. Suppose that $\Gamma(t)$ moves according to velocity $\mathbf{v}$: $\partial \Gamma(t) / \partial t = \mathbf{v}\big(\Gamma(t)\big)$, where $\mathbf{v}$ is known after the fluid dynamics equations are solved. Given the interface velocity, if we want the level set function ($d$) to satisfy $\Gamma(t) = \{\mathbf{y} : d(\mathbf{y}, t) = 0\}$ after updating, we can impose the following equation upon $d(\mathbf{y}, t)$ [97-99]

$$
\frac{\partial d}{\partial t} + \mathbf{v} \cdot \nabla d = 0
$$

$$(32)$$

with the initial condition $\Gamma(0) = \{\mathbf{y} : d(\mathbf{y}, 0) = 0\}$, i.e. the level set function initially set for the initial configuration in our case. Here the level set function was initially set as a signed distance function to the solvent accessible surface with a specified solvent probe.

### 3.3 Overview of the FD/MD numerical procedure

In our system, the atomic details for the solute region are preserved, and the solvent region is modeled as in 3.1. To simulate the solute particle dynamics, a standard MD engine with the leapfrog time integrator [100] coupled to a heat bath is used. The temperature coupling is realized with the Berendsen thermostat, which has been widely used in molecular simulation community [101]. Once the heat bath is specified, the procedure of the FD/MD can be summarized into the following steps:

```
1. Input and initialize system parameters for solute atoms
such as temperature, number of particles, time step, etc.
Initialize initial positions and velocities of all solute
atoms;
2. Initialize FD simulation box and grid points. Initialize
velocity and pressure of fluid elements;
3. Compute energy and forces from the potential function of
solute atoms;
4. Compute van del Waals forces and pressure between solute
atoms and fluid atoms;
5. Use the particle MD engine to update new velocities and
positions of solute atoms;
6. Use the FD engine to update new velocities and pressures
of fluid elements;
7. Update new FD/MD interface;
```

```
8. Repeat steps 3-7.
```

Dynamics variables, such as position, velocity, pressure, and level set function, are periodically stored after step 7 as requested. These can be used as input to restart the FD/MD simulation as needed.

## 4. Other Computational Details

For the FD simulations, physical parameters of water are set as those at 300K with viscosity $\mu = 8.51 \times 10^{-4} \, \text{Pa} \cdot \text{s}$, density $\rho = 1.00 \times 10^{3} \, \text{kg/m}^{3}$, and hydrophobic surface tension $\gamma = 8.94 \times 10^{-2} \, \text{kcal/mol} \cdot \text{A}^{2}$, with the later optimized for biomolecules given the SAS molecular surface definition in a previous work for the Amber force fields [84]. The water probe was set as 1.0 Å to set up the initial SAS surface. In the FD simulation programs, both water viscosity and density are often set as 1.0 in the internal unit. Thus proper interface between FD and MD simulation portions of our model require careful unit conversion. The details in deriving these conversion factors are given in Appendix, and the actual conversion factors are listed in Table I.

| Variable | MD Unit | FD Unit |
|---|---|---|
| Time ($t$) | 1 ps | 85.1 |
| Density ($\rho$) | 1 kg | $1.00 \times 10^{27}$ |
| Energy ($E$) | 1 kcal/mol | $9.60 \times 10^{-2}$ |

**Table I**. Conversion factors between FD and MD engines.

The FD/MD multi-scale simulation engine was developed in a revised Amber 16 release [63-65]. The Amber *ff14* force field is used to generate the topology files and the TIP3P water model is used to model the water molecules. All atomic charges were set to be zero to focus on the nonelectrostatic interactions in this study. The simulations were conducted with bonds involving hydrogen constrained. Time step was set to be 0.002 ps for both fluid dynamics region and molecule dynamics region. The temperature coupling constant is 0.2 ps in the Berendsen's thermostat to couple the temperature of the MD region, which is set to be 5 K to study the relaxation of the solute-solvent interface in this study.

Since the goal of the current development is to evaluate how well the MD/FD method reproduce the solvent interface, the MD region are restrained to focus on the FD simulation. Given that the external forces on the FD region are only van der Waals force and hydrophobic force, and they should be balanced each other at equilibrium. To speed up the relaxation, we explored both artificially increase the external force terms (by a factor of 10) or decrease the viscosity terms (by a factor of 10) to accelerate the relaxation towards equilibrium. It was found that the low viscosity runs did not relax as fast as the high force runs (data not shown). Nevertheless all alternatives will be further explored in a future study.

A single ion ($Na^+$), a single molecule n-methyl amine (NMA), and four typical small molecular complexes, adenine-thymine (AT), guanine-cytosine (GC), arginine-aspartic acid (RD) and lysine-aspartic acid (KD) were chosen to analyze the solute-solvent surface produced by the FD/MD method. In this stage of our development, the electrostatic interactions were turned off so only van der Waals and hydrophobic interactions of the solute molecules were considered though water molecules were not alternated. As a benchmark to evaluate the quality of the new multi-scale model, we conducted all-atom molecular mechanics simulations for the four tested dimer complexes to sample the solvent interface with explicit TIP3P water molecules. In these simulations, all molecules first underwent a 10,000-step energy minimization starting with a 5,000-step steepest descent followed by a 5,000-step conjugate gradient minimization. Then all solute atoms were restrained with a harmonic force constant of 50 kcal/mol-Å in all subsequent heating, equilibration, and production simulations. The molecular dynamics simulations were first heat up from 0K to 300K in 20 ps. This was then followed with a 10 ns simulation at the constant temperature of 300K and the constant pressure of 1 bar with the Berendsen heat and pressure baths. The water molecules sampled in the last 5 ns was used to analyze the solute-solvent surfaces.

## 5. Results and Discussion

### 5.1 Single atom relaxation: Reproduction of analytical solution

We first validated the FD/MD engine with a simple system with analytical solution: the solute-solvent interface of a single atom, given that the balance of hydrophobic force and van del Waals force would lead to a final equilibrium surface, a sphere with radius of $r_0$. The equilibrium can be analytically solved once the solvation free energy of the system is given as

$$G = \rho \int_{r_0}^{+\infty} \left( \frac{A}{r^{12}} - \frac{B}{r^6} \right) 4\pi r^2 \, dr + \gamma 4\pi r^2$$

$$= 4\pi\rho \left( \frac{A}{9r^9} - \frac{B}{3r^3} + \gamma r^2 \right) \tag{33}$$

Starting from a given initial state, it is expected that the system converge to its free energy minimum if there is no energy barrier, which is the case here.

In this test, an Amber sodium ion solvated in TIP3P water was used as an illustration. With the specified surface tension and van der Waals parameters from Amber 14 force field, the gradient of the free energy can be expressed as

$$\frac{\partial G}{\partial r} = 4\pi\rho(-\frac{A}{r^{10}} + \frac{B}{r^4} + 2\gamma r) \tag{34}$$

Given the values of $A = 4127 \text{ kcal/mol} \cdot \text{A}^{12}$, $B = 3.570 \text{ kcal/mol} \cdot \text{A}^6$, and $\gamma = 8.94 \times 10^{-2} \text{ kcal/mol} \cdot \text{A}^2$, the numerical solution shows that there is only one root for $\frac{\partial G}{\partial r} = 0$ when $r$ is positive, which gives the radius of the sphere to be 2.45 Å. It is also clear that $\dfrac{\partial G}{\partial r} < 0$ when $r$ approaches $0^+$ and $\dfrac{\partial G}{\partial r} > 0$ when $r$ approaches infinity. Given that 1) the gradient changes from negative to positive as $r$ changes from $0^+$ to $+\infty$, and 2) there is only one root for the gradient, it can be concluded that the gradient is negative when $r < 2.45$ Å, and positive when $r > 2.45$ Å. Thus free energy G is monotonically decreasing when $r < 2.45$ Å, and monotonically increasing when $r > 2.45$ Å. This analysis shows that there is no energy barrier in the physically allowed range of $r$.

Therefore it is possible to use a simple steepest descent minimization or a low-temperature MD relaxation to reach the global minimum in the solvation free energy. Figure I plots the evolution of volume versus time for the tested low-temperature relaxation run. It is apparent that the volume of the solute-solvent interface quickly converges to a constant volume, consistent with our analysis above. The numerical volume agrees with the analytical solution with an error of ~0.3%. Note also that the equilibrium volume is a spherical sphere for the single ion as expected.

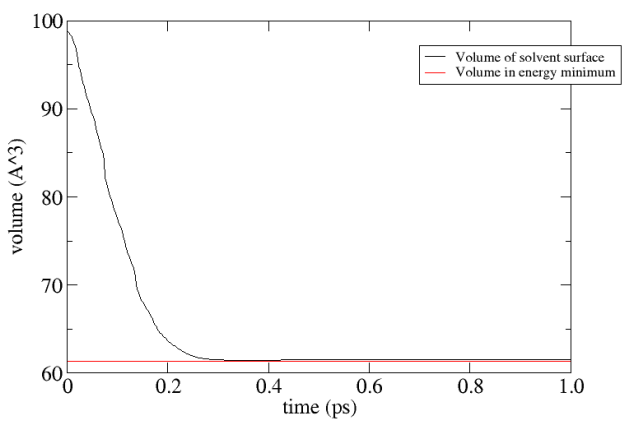

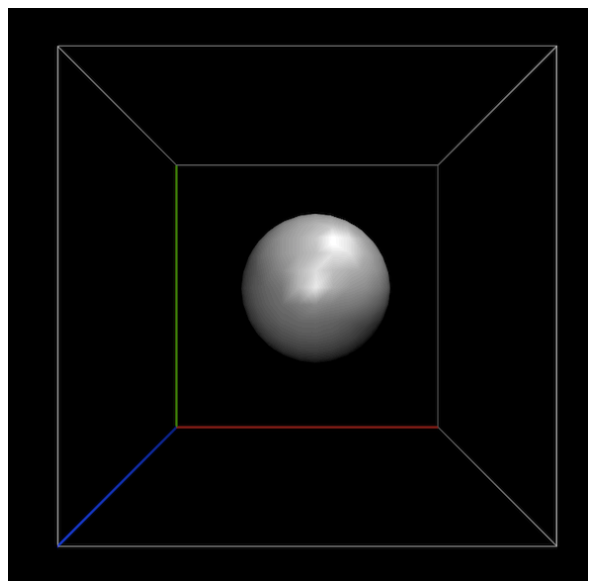

**Figure I**. (1) Time evolution of volume ($\text{Å}^3$) in the restrained FD/MD simulation of sodium ion. (2) Spherical contour of solute-solvent interface when reaching the equilibrium.

## 5.2 Monomer relaxation: Symmetric interface

Next, we performed the low-temperature relaxation of NMA, a mirror-symmetrical monomer. As shown in Figure II, the volume reaches the equilibrium value within 500 steps (1.0 ps). The contour plot shows the symmetrical monomer possesses a symmetrical interface at equilibrium. VMD visualization in 3D indicates that a detailed surface contour similar to that of the solvent excluded surface can be found (see supplementary materials).

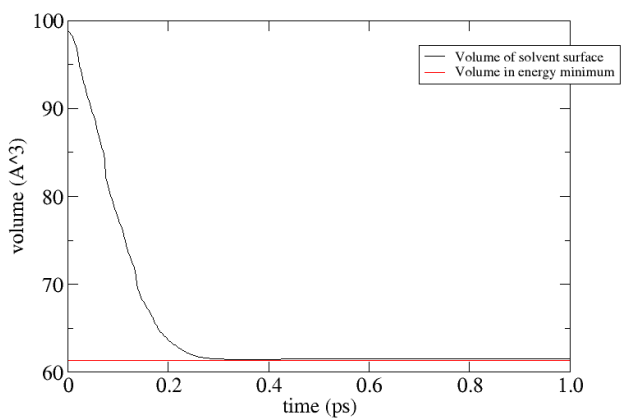

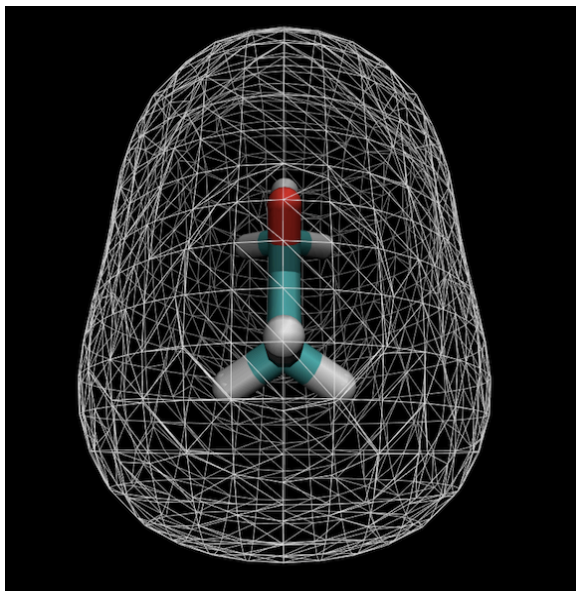

**Figure II**. (1) Time evolution of volume ($Å^3$) in the restrained FD/MD simulation of NMA. (2) SES-like solute-solvent interface is observed when reaching the equilibrium.

### 5.3 Dimer relaxation

Four typical small molecular complexes, adenine-thymine (AT), guanine-cytosine (GC), arginine-aspartic acid (RD) and lysine-aspartic acid (KD) were tested to evaluate the performance of the FD/MD simulation method. As shown in Figure III, the solute volumes reach the equilibrium values within 500 steps (1.0 ps) for all four dimers. Figure IV presents the time evolutions of force balancing on the solute-solvent interface. It is clear that the numerical solvent pressure and viscosity pressure decrease significantly and approach zero as time goes on. On the other hand, the hydrophobic (surface tension) pressure and the analytical van der Waals pressure become the dominant components, reaching steady values while balancing each other out. This is another evidence that the system approach equilibrium. Apparently the balance between hydrophobic and van der Waals components is not perfect, due to the presence of residual fluid flow nearby the solute. This issue will be addressed in our future refinement of the numerical algorithm to be discussed below.

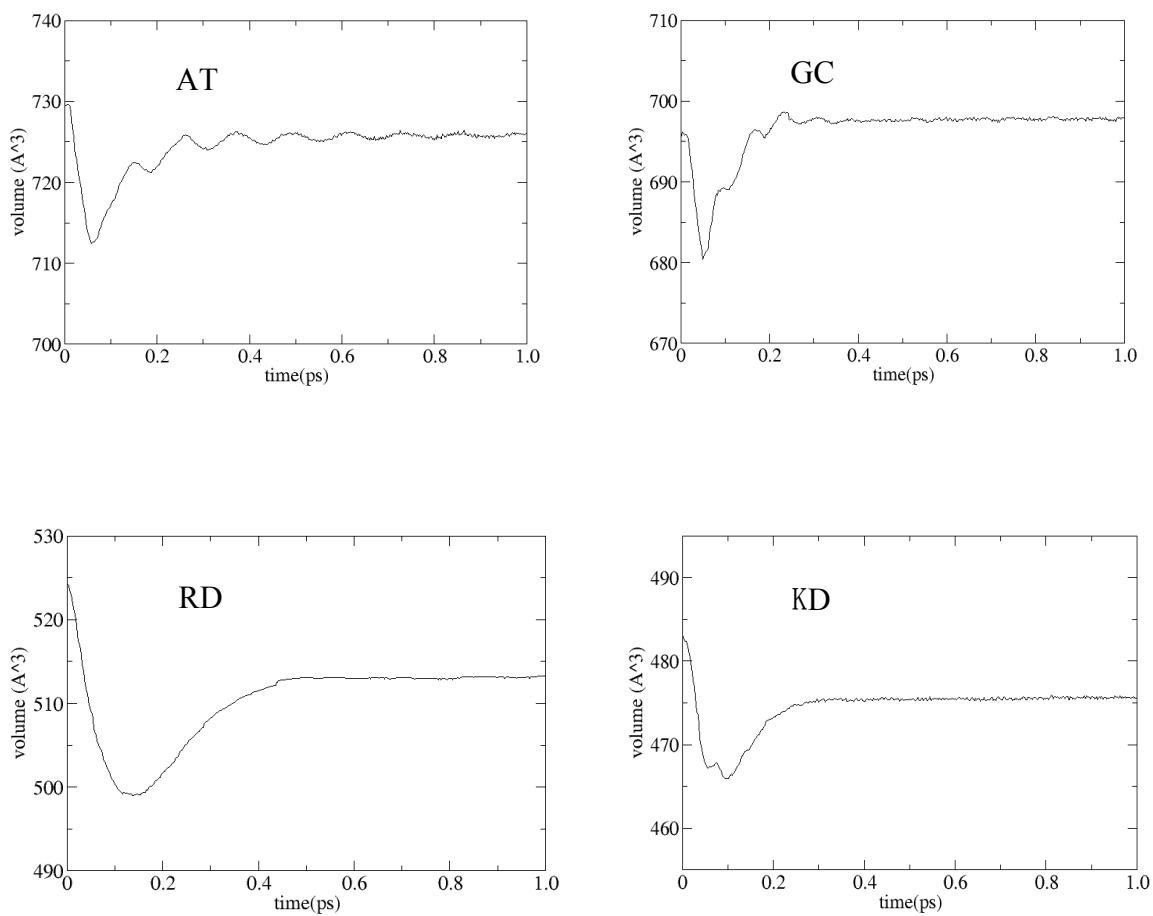

**Figure III**. Time evolutions of volume (Å³) in the restrained FD/MD simulations of dimers: adenine-thymine (AT), guanine-cytosine (GC), arginine-aspartic acid (RD) and lysine-aspartic acid (KD).

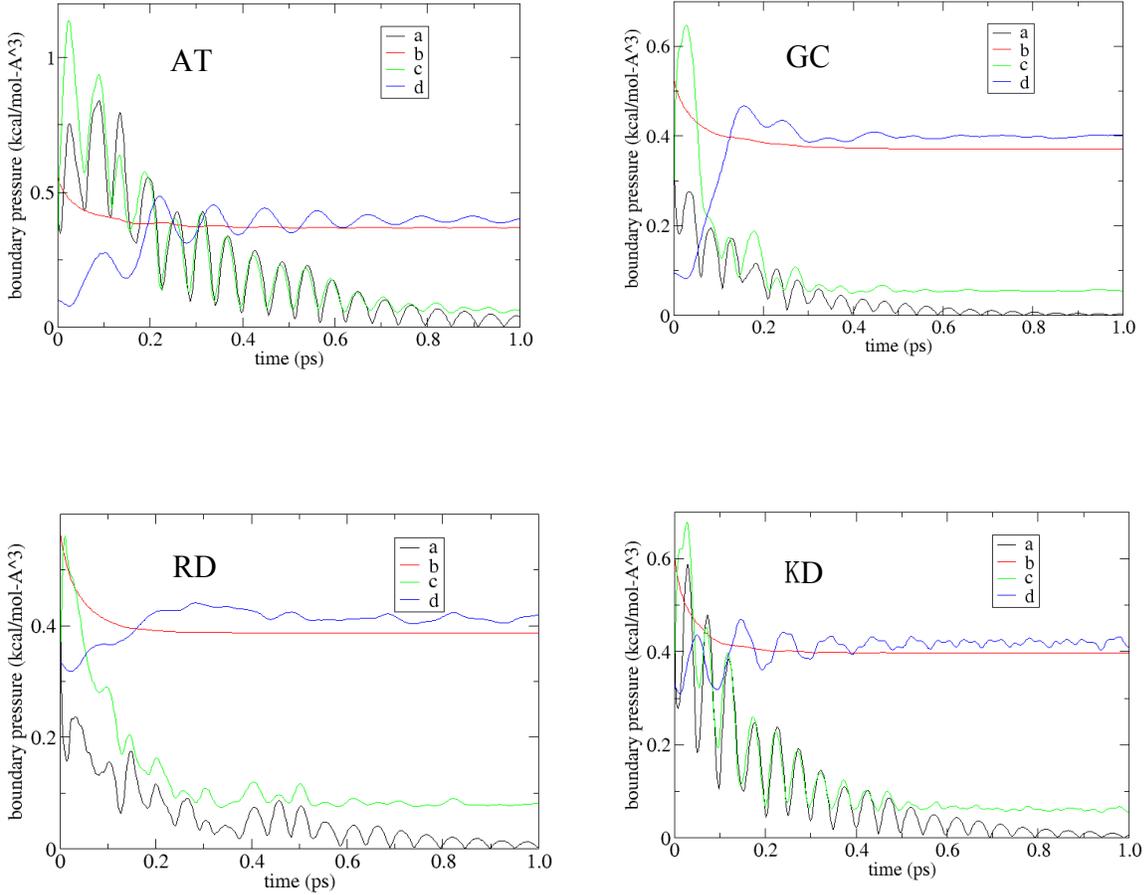

**Figure IV**. Time evolutions of average absolute pressure components on the solute-solvent interface: adenine-thymine (AT), guanine-cytosine (GC), arginine-aspartic acid (RD) and lysine-aspartic acid (KD). (a) Numerical solvent pressure; (b) Hydrophobic pressure; (c) Viscosity pressure; (d) Analytical van der Waals pressure.

## 5.4 Comparison with explicit solvent simulations

Finally, a key issue in the current development of the FD/MD model is to see whether the model at least qualitatively agrees with explicit solvent MD simulations. Discrepancy is possible given that no optimization has been attempted. Therefore, it is interesting to analyze the solute-solvent interfaces as sampled by both the FD/MD model and the explicit solvent MD model.

This analysis was conducted in the following manner. The water molecules in explicit solvent MD simulations were sampled every 5 ps over the course of a 5 ns production run for each tested dimer with all solute atoms restrained in the initial position. A total of 1000 snapshots were collected for visualization. To facilitate visualization, water molecules beyond 3.0 Å distance from any solute atom were discarded. The water distribution maps were used as references to assess the solute-solvent surface sampled by the FD/MD simulation method. Figure V shows the distribution of water oxygen atoms and the FD/MD surface when viewed outside of the solute-solvent surface, and Figure VI shows the distribution and surface when viewed inside of the solute-solvent surface. Overall, the FD/MD surfaces match very well with the solute-solvent boundaries as sampled in the explicit MD simulations for all four tested complexes. Note too there are a few places of discrepancies, which indicate that the parameters used in the FD/MD model needs to be optimized. VMD visualization in 3D further illustrates the agreement presented here (see supplementary materials).

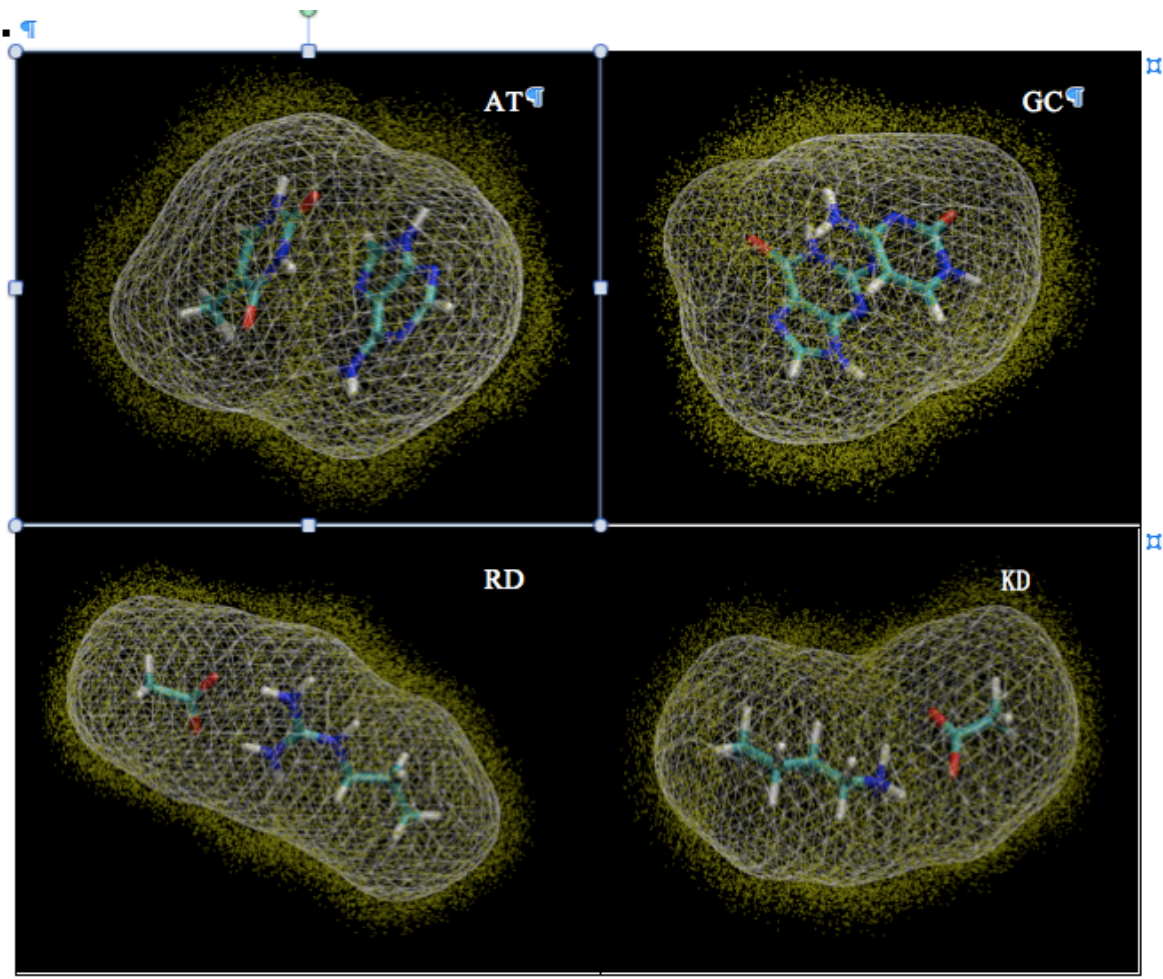

**Figure V**. FD/MD surfaces (white wireframe) and water molecules from explicit MD simulations (yellow dots) of four tested dimers: adenine-thymine (AT), guanine-cytosine (GC), arginine-aspartic acid (RD) and lysine-aspartic acid (KD). Here viewer stands outside of the surfaces.

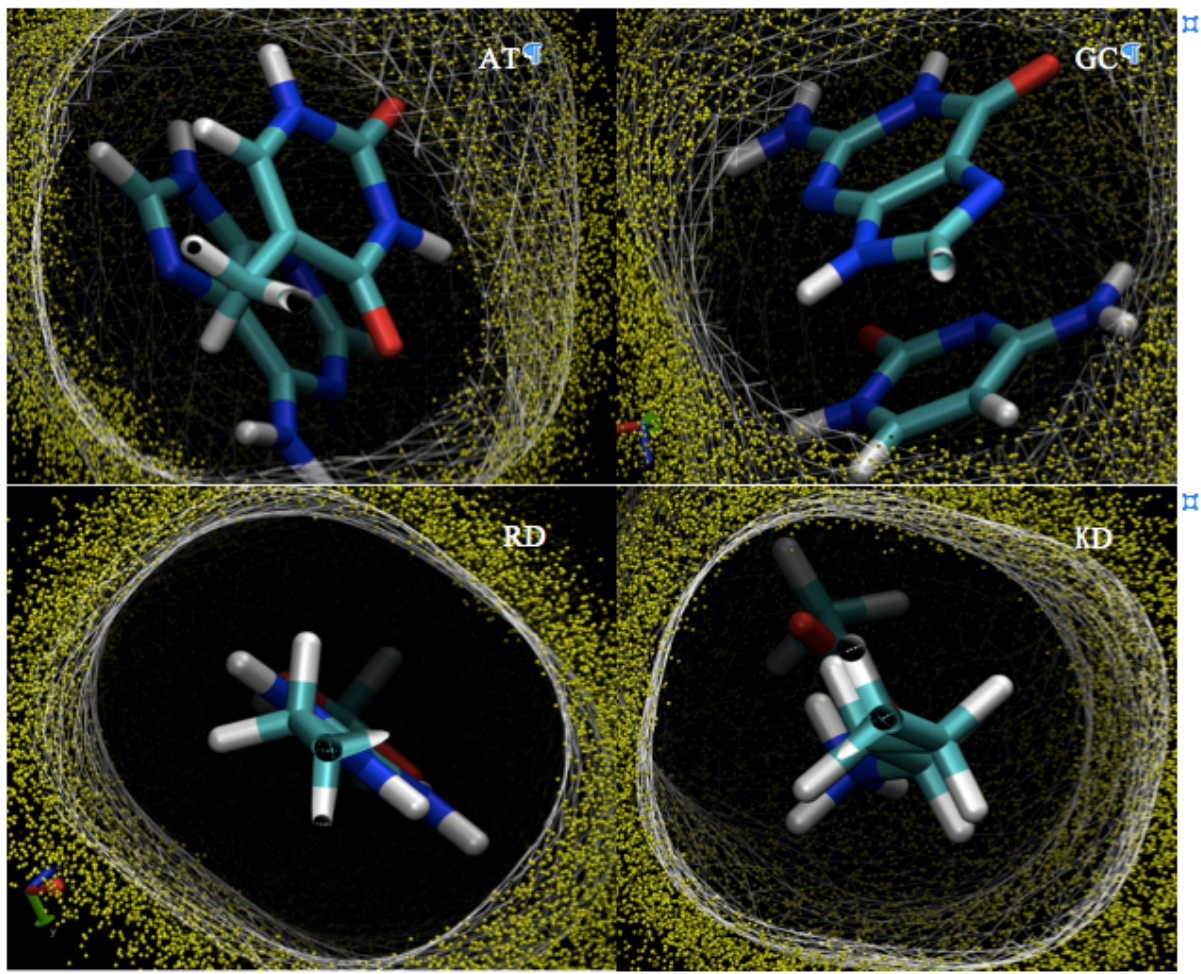

**Figure VI**. FD/MD surfaces (white wireframe) and water molecules from explicit MD simulations (yellow dots) of four tested dimers: adenine-thymine (AT), guanine-cytosine (GC), arginine-aspartic acid (RD) and lysine-aspartic acid (KD). Here viewer stands inside of the surfaces.

## 5.5 Limitations of the model and future directions

There are clearly limitations in the proposed FD/MD model. The first limitation is that we artificially make both hydrophobic and van der Waals term 10 times higher to accelerate the solute-solvent interface relaxation because the focus of the current model is for equilibrium properties of the solute, but not the physically correct solvation relaxation process, which may be important if the FD/MD model is applied to study hydrodynamic properties due to the presence of the molecular solute. Nevertheless, the artificial setting does not affect the converged solute-solvent interface because both hydrophobic and van der Waals pressure are simultaneously increased. Secondly, the finite-difference grid spacing used in the FD engine is 0.5 Å, which is widely used in biomolecular applications of a finite-difference method given a high enough resolution of molecular surface topology can be achieved. However, the relatively fine grid also leads to highly inefficient numerical procedure. To date we have not paid special attention to the numerical efficiency of our FD engine, and this will be a focus in our future development. The development and illustrations here mainly show that the FD/MD model is sound and it does produce physically meaningful observations consistent with the all-atom MD model, which is very promising.

As we pointed out in Other Computational Details the FD parameters for the water solvation process was from a previous study to optimize a related nonpolar solvent model. Apparently this is not optimal for the current FD/MD model. Our next step will be to investigate how to optimize the hydrophobic term and van der Waals term to best reproduce the all-atom explicit solvent MD model. In addition, we will also incorporate the electrostatic interaction as modeled by the Poisson-Boltzmann method to build a more realistic FD/MD model for biomolecular applications. To study fluid dynamic properties due to the presence of molecular solutes, we think the best strategy is to incorporate a coarse-grained molecular model instead of the all-atom model to make it a viable approach for systems with interesting hydrodynamic properties.

## 6. Conclusions

A multi-scale framework was recently proposed for more realistic molecular dynamics simulations in continuum solvent models by coupling a molecular mechanics treatment of solute with a fluid mechanics treatment of solvent [61]. Our previous work addressed the mathematical issues in solving fluid dynamics equations numerically [61, 62]. In this study we incorporated the fluid dynamics algorithm with the Amber molecular mechanics package [63-65] to conduct atomistic simulations of biomolecules. A major issue in the application of the multi-scale model in atomistic simulations is the presence of van del Waals potential, which has a large gradient nearby the solute-solvent interface. It is virtually impossible to treat van der Waals potentials with any reasonably fine finite-difference method. We overcame the challenge by removing the van del Waals potential from pressure when solving the finite-difference fluid dynamics equations, and adding back the van del Waals potential analytically in the free-boundary condition.

We first validated the FD/MD engine with a simple system with analytical solution: the solute-solvent interface of a single atom. The balance of hydrophobic force and van del Waals force would lead to the final equilibrium surface of a sphere. Our test shows that the volume of the solute-solvent interface quickly converges to the analytical value with an error ~0.3%. Next, we performed the relaxation of NMA, a mirror-symmetrical monomer. The contour plot shows the symmetrical monomer possesses a symmetrical interface at equilibration. VMD visualization in 3D indicates that a detailed surface contour similar to that of the solvent excluded surface can be found. Four typical small molecular complexes were then tested to evaluate the performance of the FD/MD simulation method. The solute volumes reach the equilibrium values within 1.0 ps for all four dimers. The time evolutions of force balancing analysis on the solute-solvent interface show that the numerical solvent pressure and viscosity pressure decrease significantly and approach zero as simulation time goes on. On the other hand, the hydrophobic (surface tension) pressure and the analytical van der Waals pressure become the dominant components, reaching steady values while balancing each other out. This strongly indicates that the systems approach the equilibrium at the end of the simulations.

Finally, a key issue at the current stage of the development is to investigate whether the model at least qualitatively agrees with explicit solvent MD simulations. Therefore, it is interesting to analyze the solute-solvent interfaces as sampled by both the FD/MD model and the explicit solvent MD model. Comparisons show that the FD/MD surfaces agree very well with the solute-solvent boundaries as sampled in the explicit MD simulations for all four tested dimers. Note too a few places of discrepancies do exist, which indicate that the parameters used in the FD/MD model needs to be optimized further to achieve higher consistency with the all-atom explicit solvent MD model.

In our next phase of the development, we will further improve the quality of hydrophobic and van der Waals terms of the FD model to best reproduce all-atom force field model. It is also interesting to investigate the effect of incorporating the electrostatic forces into the FD/MD model to evaluate its impact on both numerical stability and consistency for a range of model systems. Finally it is also interesting to explore more efficient and more robust numerical FD engines for routine applications to biomolecular systems.

## Supplementary Materials

Supplementary materials are available online for the 3D visualization in VMD for monomer NMA, and the four molecular dimer complexes: adenine-thymine (AT), guanine-cytosine (GC), arginine-aspartic acid (RD) and lysine-aspartic acid (KD).

## Acknowledgements

This work was supported in part by NIH/NIGMS (GM093040 & GM079383).

## Appendix

Since FD programs can use any arbitrary length unit, we set 1 internal length unit as 1 Ångstrom. Given the length unit settled and the water density ($\rho = 1.00 \times 10^3 \text{kg/m}^3$) set as 1 internal density unit, the internal mass unit can be computed to be equivalent to $1.00 \times 10^{-27} \text{kg}$. Next we can utilize viscosity to compute the time conversion factor. Given the unit of viscosity $\text{Pa} \cdot \text{s} = \text{kg/(m} \cdot \text{s)}$ we can use the mass conversion factor to derive the time conversion factor as follows

$$\begin{aligned} 1 \text{ internal viscosity unit} &= 8.509 \times 10^{-4} \text{kg/(m} \bullet \text{s)} \\ &= 8.509 \times 10^{-4} \times 1.00 \times 10^{27} / (10^{10} \times T) \end{aligned} \quad \text{A.1}$$

This leads to $T = 8.51 \times 10^{13}$, which means $1 \, s = 8.51 \times 10^{13}$ internal time unit. And thus we have $1 \, ps = 10^{-12} \, s = 85.1$ internal time unit. The energy unit of 1 kcal/mol can be converted as

$$\begin{aligned} 1 \text{ kcal/mol} &= 6.948 \times 10^{-21} \text{ J} = 6.948 \times 10^{-21} \text{ kg m}^2/\text{s}^2 \\ &= 6.948 \times 10^{-21} \times 10^{27} \times (10^{10})^2 / (8.509 \times 10^{13})^2 \text{ internal energy unit} \\ &= 9.60 \times 10^{-2} \text{ internal energy unit} \end{aligned} \quad \text{A.2}$$